\documentclass[aps,showpacs,preprint,superscriptaddress,12pt]{revtex4}
\usepackage{graphicx}
\usepackage{bm}
\usepackage{txfonts}

\begin{document}
\title{The effect of frequency chirping on electron-positron pair production in the one- and two-color laser pulse fields}

\author{Nuriman Abdukerim}
\affiliation{College of Nuclear Science and Technology, Beijing Normal University, Beijing 100875, China}
\author{Zi-Liang Li}
\affiliation{College of Nuclear Science and Technology, Beijing Normal University, Beijing 100875, China}
\author{Bai-Song Xie  \footnote{Corresponding author. Email address: bsxie@bnu.edu.cn}}
\affiliation{College of Nuclear Science and Technology, Beijing Normal University, Beijing 100875, China}
\affiliation{Beijing Radiation Center, Beijing 100875, China}
\date{\today}
\begin{abstract}
The effect of the frequency chirping on momentum spectrum and pair production rate in one- and two-color laser pulse fields is investigated by solving the quantum Vlasov equation. A small frequency chirp shifts the momentum spectrum along the momentum axis. The positive and negative frequency chirp parameters play the same role in increasing the pair number density. The sign change of frequency chirp parameter at the moment $t=0$ leads pulse shape and momentum spectrum to be symmetric, and the number density to be increased. The number density of produced pairs in the two-color pulse field is much higher than that in the one-color pulse field and the larger frequency chirp pulse field dominates more strongly. In the two-color pulse fields, the relation between the frequency ratio of two colors and the number density is not sensitive to the parameters of small frequency chirp added in either low frequency strong field or high frequency weak field but sensitive to the parameters of large frequency chirp added in high frequency weak field.

\end{abstract}

\maketitle

\section{Introduction}

The vacuum pair production in the external electric field in the frame of quantum electrodynamics (QED) was first proposed in 1931 by Sauter \cite{Sauter} and also studied in 1936 by Heisenberg \cite{Heisenberg}. In particular in 1951, Schwinger studied this problem systemically by employing the proper-time technique \cite{Schwinger}. He got the critical value of the constant external field to create the electron-positron  ($e^{-}e^{+}$) pairs from the vacuum as $E_{cr}=m_{e}^{2}c^3/e\hbar \approx 1.3\times 10^{16}\rm{V/cm}$, where $m_{e}$ is the electron mass and $-e$ is an elementary electron charge. This critical field strength is related to the laser intensity $I\approx4.3\times 10^{29}\rm{W/cm^{2}}$. The intensity is so high really, however, as the new experimental development in ultrahigh intensity laser techniques \cite{Ringwald, XFEL,Tajima, ELI}, it may be possible to get subcritical laser field \cite{Roberts, Alkofer}.

On the other hand, because the propose of dynamically assisted Schwinger mechanism, which combines different pulse laser fields,
the vacuum pair creation may be observed even in much lower intensity laser field \cite{Schutzhold, Dunne1, Bell, Bulanov1, Monin, Piazza}. So many ways can be used to reduce the requirement of laser field but enhance the pair production, for example, the combination of sinusoid/ cosine with exponential laser pulse \cite{Hebenstreit1, Nuriman1, Obulkasim}, by using super-Gaussian instead of Gaussian pulse to widen pulse width \cite{Nuriman2,Dumlu2}, the usages of multi-slit interference or/and the alternating sign $N$-pulse electric fields \cite{Li2, Li3} and so on.

Beside changing the shape or carrier phase \cite{Hebenstreit1,Nuriman2}, it is also expected and found that, through adding  a small frequency chirping, it can influence the momentum distribution and then possibly increase the created  $e^{-}e^{+}$ pair number density \cite{Dumlu1, Jiang}. In \cite{Jiang}, the field time was divide into three parts and different frequency chirps are applied. In\cite{Dumlu1}, different frequency chirps lead the changing of the momentum spectra and it is explained by turning point.
These fewer research make this topic very interesting meanwhile make some problems still open. For example, whether the momentum spectra and pair number are dependent on the sign of
frequency chirping? What phenomena occur if we change the one-color pulse with single frequency chirping to two-color pulse with double frequencies chirping or if we change the sign of chirp parameter from positive (negative) to negative (positive) at the peak position of one-color laser pulse? How the two-color laser pulse field with different chirping  affect the momentum spectra and the pair production rate?

In this paper, we will answer these questions. We have studied the one-color and two-color laser pulse fields by adding a positive or negative frequency chirp parameters and analyze the effects of the frequency chirp parameters on the laser pulse shape, momentum spectrum and number density. It is found that a small frequency chirp can shift the momentum spectrum along the momentum axis. The positive and negative frequency chirp parameters play the same role in increasing the pair number density. The sign change of frequency chirp parameter at the moment $t=0$ leads pulse shape and momentum spectrum to be symmetric. The case of first positive and then negative frequency chirp lead to the wider time lasting of pulse within the neighbor of peak field strength at $t=0$ than the opposite case of first negative and then positive chirp, therefore, the pair number density dominated by tunneling mechanism is larger in the former case. On the other hand, the number density of produced pairs in the two-color pulse fields is also much higher than that in the one-color pulse field. Moreover, in two-color case, the relation between the number density and the frequency ratio of two colors is not sensitive to the small frequency chirp field but sensitive to the large frequency chirp field.

\section{Theoretical formalism based on quantum Vlasov equation}

For the completeness of the paper, we have to include the description about the basic ideas and formula of quantum Vlasov equation while the content would be unavoidable repetition of other published papers. Here we outline it similar to Ref.\cite{Nuriman2}.

The source term of pair production, $s(\mathbf{p}, t)$, obviously depends on the applied
external field as well as the electron/positron kinetic property.
From $df(\mathbf{p}, t)/dt = s(\mathbf{p}, t)$, where $f(\mathbf{p}, t)$ is the momentum distribution of the created pairs, we get the quantum Vlasov equation (QVE) in the
following integro-differential equation form
\begin{equation}
\frac{df(\mathbf{p}, t)}{dt} =
\frac{eE(t)\varepsilon_{\perp}^{2}}{2\omega^{2}(\mathbf{p}, t)}
\int_{t_0}^{t}dt'\frac{eE(t')[1-2f(\mathbf{p}, t')]}{\omega^{2}(\mathbf{p}, t')}
\cos\left[2\int_{t'}^{t}d\tau\omega(\mathbf{p}, \tau)\right],
\label{eq2}
\end{equation}
where the quantities are the electron/positron momentum $\mathbf{p}=(\mathbf{p}_{\bot}, p_{\parallel})$,  transverse energy-squared $\varepsilon_{\bot}^{2}=m_{e}^{2}+p_{\bot}^{2}$, the total energy-squared
$\omega^{2}(\mathbf{p}, t)=\varepsilon_{\bot}^{2}+p_{\parallel}^{2}$, and the longitudinal momentum $p_{\parallel} = P_{3}-eA(t)$. If we define $q(\mathbf{p}, t)=eE(t)\varepsilon_{\bot}/\omega^{2}(\mathbf{p}, t)$
and $\Theta(\mathbf{p}, t', t)=\int_{t'}^{t}\omega(\mathbf{p}, \tau)d\tau$ then Eq. (\ref{eq2}) becomes
\begin{equation}
\frac{df(\mathbf{p}, t)}{dt} =
\frac{1}{2}q(\mathbf{p}, t)\int_{t_0}^{t}dt' q(\mathbf{p}, t')
[1-2f(\mathbf{p}, t')]\cos[2\Theta(\mathbf{p}, t', t)].  \label{eq3}
\end{equation}
Moreover when the integral part is represented by
$g(\mathbf{p}, t)$ in Eq. (\ref{eq3}), the equation can be expressed
as a set of first order ordinary differential equations (ODEs)
\cite{Hebenstreit1}
\begin{eqnarray}
\dot{f}(\mathbf{p}, t) & = & \frac{1}{2}q(\mathbf{p}, t)g(\mathbf{p}, t), \\
\dot{g}(\mathbf{p}, t) & = & q(\mathbf{p}, t)[1-2f(\mathbf{p}, t)]
-2\omega(\mathbf{p}, t)w(\mathbf{p}, t),  \\
\dot{w}(\mathbf{p}, t) & = & 2\omega(\mathbf{p}, t)g(\mathbf{p}, t).
\label{eq4}
\end{eqnarray}
For convenience and simplicity we denote the derivative of a physical quantity with respect to time by a
symbol dot above this quantity. It should be emphasized that the change from the original integral-differential equation to a set of ODEs not only makes the numerical treatment simpler but also makes the involved physical quantities or/and terms clearer. For example the term $g(\mathbf{p},t)$, i.e. the integral part of Eq.(\ref{eq3}), constitutes an important contribution to the source of pair production. In fact this term reveals also the quantum statistics character through the term $[1-2f(\mathbf{p},t)]$ due to the Pauli exclusive principle. On the other hand, $w(\mathbf{p},t)$ denotes a countering term to pair production, which is associated to the pair annihilation in pair created process to some extent. Obviously the last one of ODEs means that the more pairs are created, the more pairs are annihilated probably in pair created process. Thus combining all factors aforementioned will conclude that the studied system exhibits a typical non-Markovian character.

The initial condition of Eq. (\ref{eq3}) can be given as $f(\mathbf{p}, t_0)$,
$g(\mathbf{p}, t_0)$ and $w(\mathbf{p}, t_0)$ in terms of concrete
physical problem. Integrating the distribution function to
momentum we can get the time-dependent $e^{-}e^{+}$  pair number
density as
\begin{equation}
n(t)=2\int\frac{d^{3}\mathbf{p}}{(2\pi)^{3}}f(\mathbf{p}, t),
\label{eq-number}
\end{equation}
which will be very useful in the following study on pair creation enhancement.
Since the applied laser field becomes zero when $t \rightarrow \infty$, therefore, what we are interested in are the stationary distribution function $f=f(\textbf{p}, t \rightarrow \infty)$ as well as the number density $n=n (t \rightarrow \infty)$.

We use the electrons quantities as the normalized ones, i.e. length $\lambda_{c}=2\pi\hbar/m_{e}c=2. 43\times10^{-12}\rm{m}$, time $\tau_e=\lambda_{c}/c=1. 287\times10^{-21} \rm{s}$ and the momentum $m_{e}c=0. 51 \rm{Mev}/c$. The field strength is normalized by $E_{cr}$. In our study some typical parameters are given as laser frequency $\omega=0.02$ or/and $\omega=0.2$, the pulse length $\tau=100$ and the field strength is $E=0.1$ or/and $E=0.01$ and so on.

\section{Numerical results of momentum distribution and pair number density}

In this section, we will analyze the effect of the frequency chirp parameter on pair production rate by discussing the laser pulse field, momentum spectrum and the number density.

\subsection{One-color laser pulse field}

The one-color laser pulse field is given by
\begin{equation}
E_1(t)=E_{0} e^{-\frac{t^2}{2\tau^2}} \cos(bt^{2}+\omega t),
\label{onefield}
\end{equation}
where $E_{0}=0.1$, $\omega=0.02$, and the pulse width $\tau=100$, $b$ is the frequency chirp parameter. After adding a small frequency chirp, the laser frequency becomes a linear variation with time as $\omega'=(bt+\omega)$, can be called effective frequency. By the way in general the frequency chirping in the all pulse time should not beyond the original frequency $\omega$ so that $b<\omega/\tau$ is required. For simplicity we do not consider the transverse momentum, i.e., $p_{\perp}=0$.

\begin{figure}
\begin{center}
\includegraphics*[width=16cm,keepaspectratio]{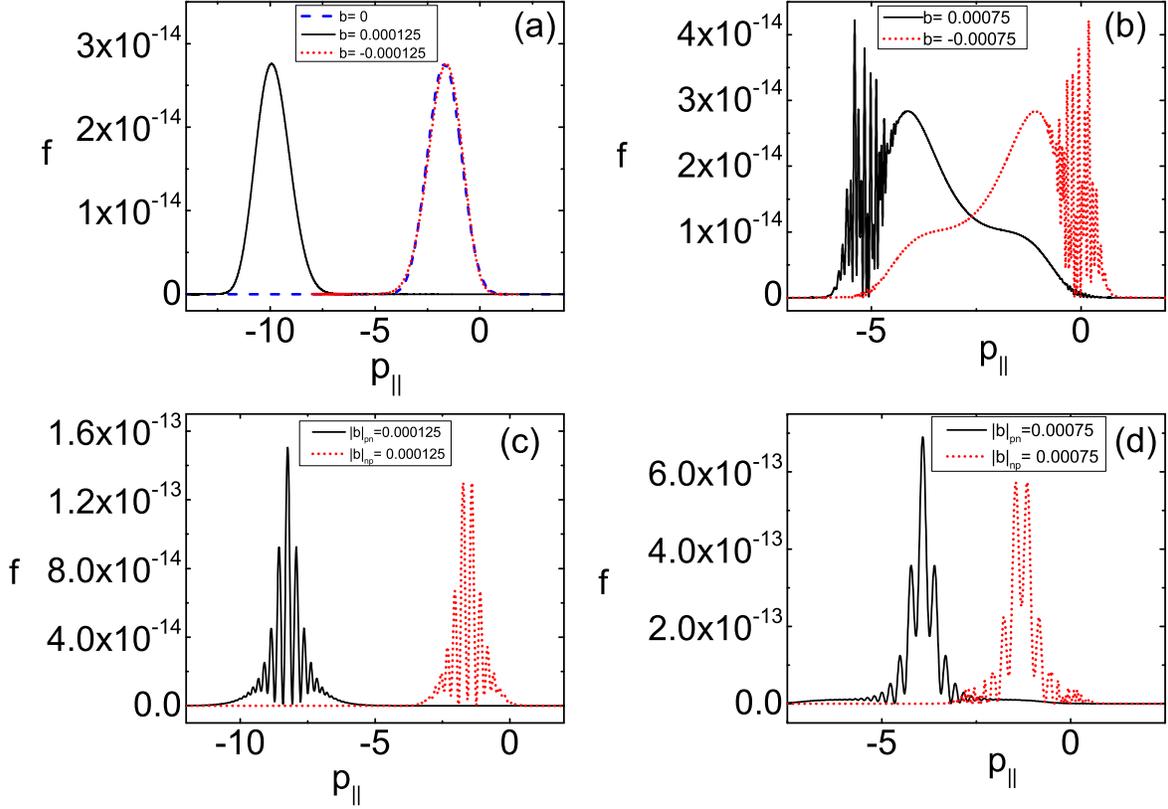}
\caption{\label{Fig. 1} (color online) Momentum spectrum of the longitudinal momentum $p_{\parallel}$ of the produced electron-positron pairs in one-color pulse laser field $E_1 (t)$ with different frequency parameter $b$.}
\end{center}
\end{figure}

By adding different frequency chirps on the one-color laser pulse field, the momentum spectrum is plotted in Fig.\ref{Fig. 1}. First we add a small frequency chirp with positive $b=0.000125$ and negative $b=-0.000125$ respectively. 
Compared to the momentum spectrum of free-chirp $b=0$ pulse field (blue dashed line), it is found the momentum spectrum of positive chirp (black solid line) is shifted greatly to the negative direction while the momentum spectrum of negative chirp (red dotted line) is hardly shifted, see Fig.\ref{Fig. 1}(a).

It can be also seen that when a small frequency chirp is applied the maximum value of momentum spectrum does not increase.
This momentum spectrum shifting by the small positive frequency chirp may be meaningful to the  spectrometry measurements for pairs because they are sensitive in a certain momentum window. The frequency chirp shifts the maximum value of momentum spectrum to the characteristic momentum range that can increase the detection probability, refer to \cite{Hebenstreit2016The}. By the way the carrier phase can also shift the momentum spectrum \cite{Dumlu2010Schwinger,Hebenstreit2009Momentum}, however, for periodic laser pulse, the variable carrier phase maybe leads the decreasing of the momentum spectrum sometimes \cite{Nuriman2}.

Second we add a larger positive frequency chirp $b=0.00075$ (the black solid line) or a negative frequency chirp $b=-0.00075$ (the red dotted line), as shown in Fig.\ref{Fig. 1}(b). Now we get the highly asymmetric momentum spectrum meanwhile with very irregular oscillation. And the peak of momentum spectrum is increased also. Positive frequency chirp shifts still the momentum spectrum to the negative direction while negative frequency chirp shifts the momentum spectrum to the positive direction very little still, which depends on the concrete effective frequency. The irregular oscillation may be understand as in the scattering picture as that the corresponding effective scattering potential changes by a variation of frequency chirp parameter \cite{Dumlu2010Schwinger}.

Third we try to change the chirping sign before and after the moment $t=0$. Fig.\ref{Fig. 1}(c) correspond two cases. One is that we add the small positive frequency chirp $b=0.000125$ when $t\leq0$ and then small negative frequency chirp $b=-0.000125$ when $t>0$. The other is vise versa. The momentum spectrum in the former case is plotted with black solid line and the other case is plotted with red dotted line. It is found that the sign frequency chirp causes a field pulse shape to be symmetric which leads to the momentum spectrum symmetric also, and moreover the higher number density is got. In this situation, $|b|=0.000125$, the momentum spectrum not only shifts but also it's value is larger one order of magnitude than the chirp-free one. By the way the black solid line exhibit a single peak while the red dotted line exhibit the double-peak structure. Interestingly if the sign chirping strength is increased to $b=0.00075$, results are shown in Fig.\ref{Fig. 1}(d), the symmetric structure of the momentum spectrum becomes simpler and the maximum value of them increases greatly, about four times comparable to the case of $b=0.000125$.

\begin{figure}
\begin{center}
\includegraphics*[width=12cm,keepaspectratio]{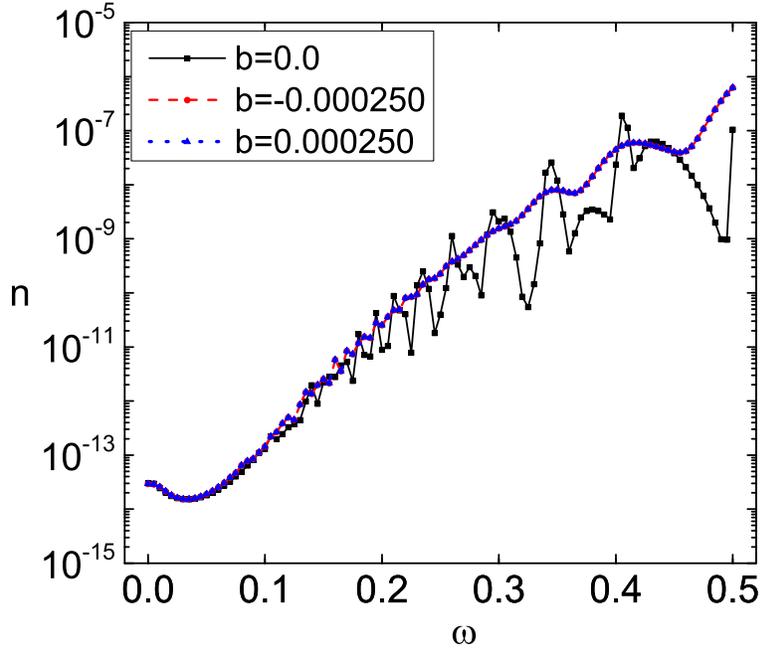}
\caption{\label{Fig. 2} (color online) Electron-positron number density vs the original frequency in one-color pulse laser field $E_1 (t)$.}
\end{center}
\end{figure}

\begin{figure}
\begin{center}
\includegraphics*[width=12cm,keepaspectratio]{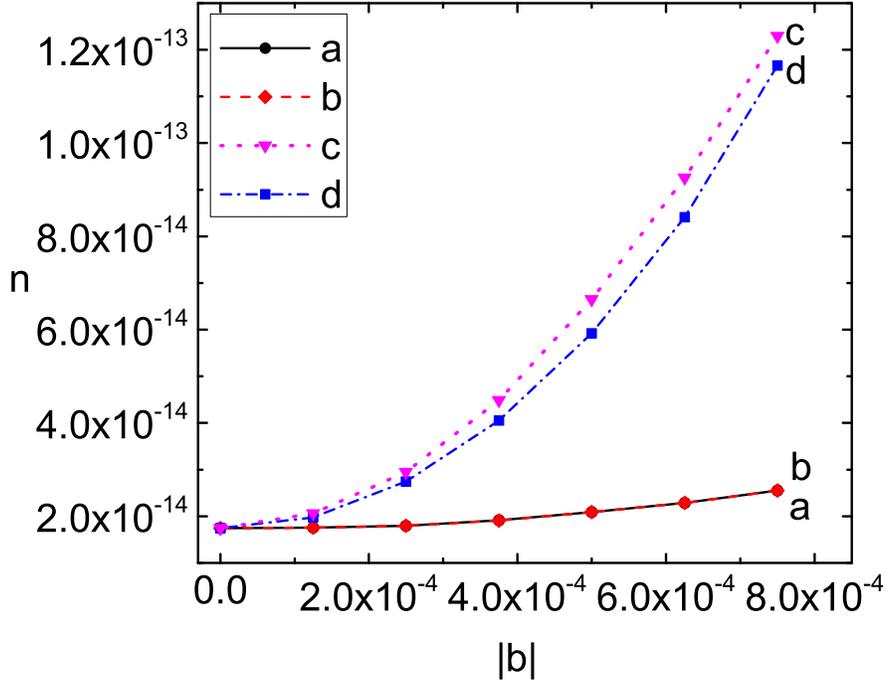}
\caption{\label{Fig. 3} (color online) Electron-positron number density vs the frequency chirp parameter $|b|$ in one-color pulse laser field $E_1 (t)$.}
\end{center}
\end{figure}

Now let us to see what happen if we keep the frequency chirp parameter fixed but change the original frequency $\omega$. The typical results are depicted in Fig.\ref{Fig. 2} in which three cases are studied by positive chip $b=0.00025$, negative chirp $b=-0.00025$ and without chirp $b=0$. The variation of the original frequency $\omega$ is from $0$ to $0.5$. It can be seen that, when $\omega<0.1$, the number density decreased first and then increased, the three curves are almost coincident. When $\omega>0.1$, as the frequency increases, the curves presented pronounced oscillation, which is similar to the situation in paper \cite{Nuriman2,Christian2014Effective} that caused by the multiphoton process. The oscillation of curve $b=0$ is more pronounced. The adding of frequency chirping results in smoothing of number density and widening of the peaks and the results of positive and negative chirp are coincide with each other. Moreover for most of $\omega$ the number density of pairs created in the chirped laser pulse is higher than that of the chirp-free one. For example, at $\omega=0.325$, the number density $n=3.585\times10^{-09}$ in the case of chirping is two orders of magnitude higher than $n=5.4\times10^{-11}$ in the chirp-free one. As $\omega$ increases, the difference becomes larger more. For example, when $\omega=0.49$, $n=9.8\times10^{-10}$ is got for chirp-free laser pulse and $n=3.5\times10^{-07}$ is got for other two chirping cases, obviously, the difference between them is almost three orders of magnitude.

For one-color field finally let us to see how the pair number density depends on the frequency chip. For fixed $\omega=0.02$ fixed, the number density vs the frequency chirp parameter is depicted in Fig.\ref{Fig. 3}.
The lines labeled $(a)$, $(b)$ are number density curves related to the positive and negative frequency chirp parameters, respectively. The line $(c)$ is number density curve related to the frequency chirp parameter which is positive when $t\leq0$ and negative when $t>0$ and   $(d)$ is the opposite case about the frequency sign . It can be seen that the curves $(a)$ and $(b)$ coincide completely. It is not surprising because when $t\rightarrow-t$ the positive chirped laser pulse field and the negative one is the same so that they have the time-reverse symmetry. As the frequency chirp parameter $|b|$ increases the number density increases also. But the number densities in cases of $(c)$ and $(d)$ are higher than those of $(a)$ and $(b)$. This can be understand from the point of turning points. The laser pulse fields related to $(c)$ and $(d)$ are symmetry compared to the $(a)$ and $(b)$ so that the more turning points exist in the complex $t$ space. Then the interference between the turning point pairs causes complex oscillation structure in the momentum which can spectrum~\cite{Hebenstreit2009Momentum,Dumlu2010Stokes,Eric2012Ramsey,Akal2014Electron} consequently increase the pair production rate \cite{Dumlu2}. On the other hand the number density of $(c)$ is higher than $(d)$ may be attributed a fact that the strong but slow varying field around time $t=0$ triggers the tunneling process of Schwinger pair creation has a little difference in two cases. Obviously this slow varying field has a wider width around time $t=0$ in case $(c)$ than $(d)$ because when $t\neq 0$ the frequency is decreasing in (c) but is increasing in (d).

\subsection{Two-color laser pulse field}

It is mentioned in some works \cite{Christian2014Effective, Li2} that as the pulse number increases the pair production rate will be enhanced. In particular by a high frequency weak field superimposed on the low frequency strong field the dynamically assisted mechanism can enhance greatly the pair production \cite{Schutzhold}. Now let us to investigate this problem under the multi-color pulse fields by including the effect of frequency chirp parameter on the pair production rate. For simplicity we only consider two-color laser pulse fields by adding the frequency chirping. The fields can be given as
\begin{equation}
E(t)=E_1(t)+E_2(t)=E_{1}e^{-\frac{t^2}{2\tau^2}} \cos(b_{1}t^{2}+\omega_{1} t)+
E_{2}e^{-\frac{t^2}{2\tau^2}} \cos(b_{2}t^{2}+\omega_{2} t),
\label{twofield}
\end{equation}
where $\omega_{1}=0.02$, $\omega_{2}=10\omega_1=0.2$, $E_{1}=0.1$, $E_{2}=0.1 E_2=0.01$ and $\tau=100$. The frequency chirp parameters are $b_1$ and $b_2$ for low frequency field $E_1(t)$ and high frequency field $E_2(t)$, respectively, and in general $b_1<\omega_1/\tau$, $b_2<\omega_2/\tau$ are required.

\begin{figure}
\begin{center}
\includegraphics*[width=16cm,keepaspectratio]{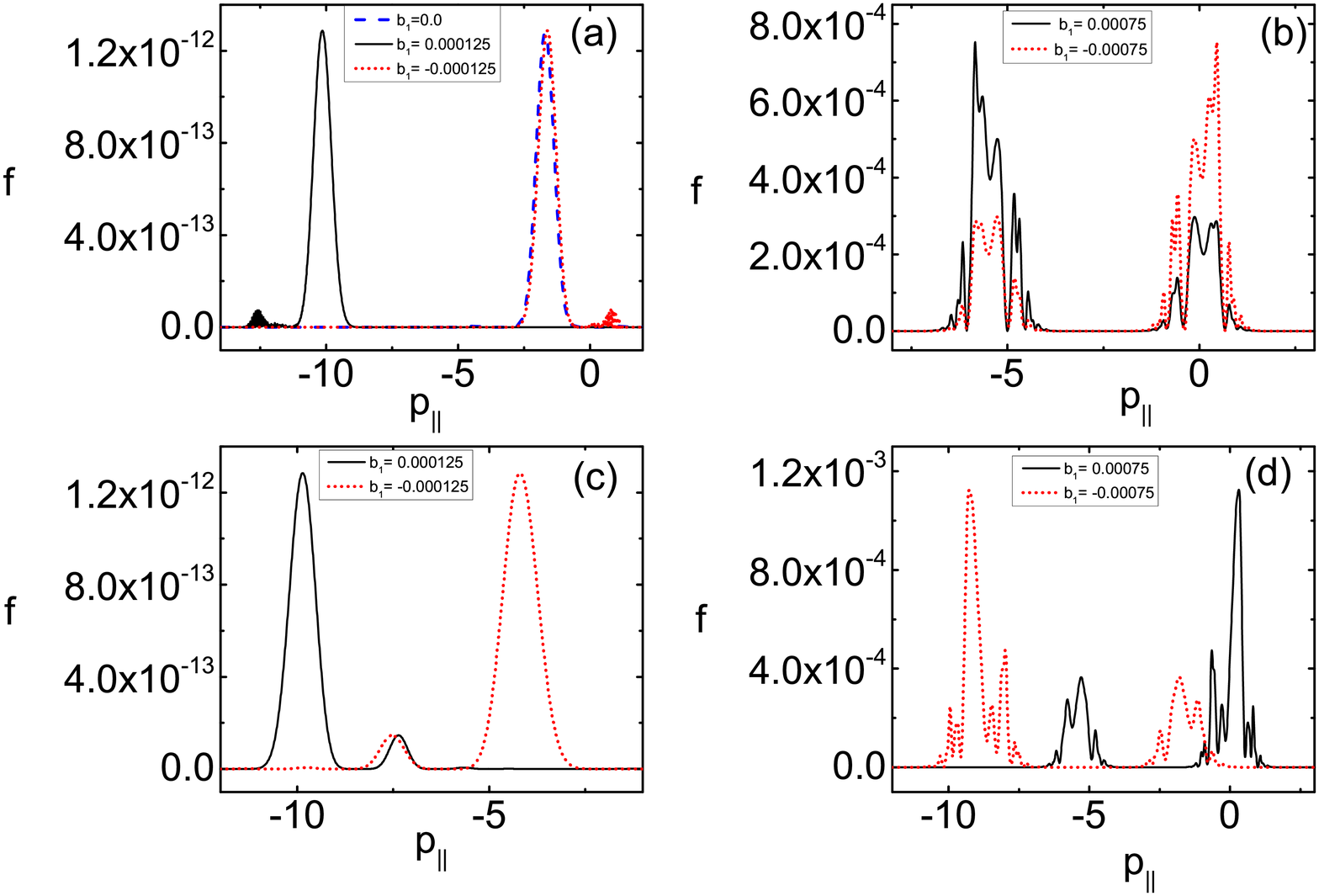}
\caption{\label{Fig. 4} (color online) The same as in Fig.\ref{Fig. 1} except in two-color pulse laser field $E_{1}(t)+
E_{2}(t)$ with different frequency parameters $b_1$ and $b_2$.}
\end{center}
\end{figure}

For some typical frequency chirp parameters $b_{1}$ and $b_{2}$ the momentum spectrum are depicted in Fig.\ref{Fig. 4}. By comparing the blue dashed lines in Fig.\ref{Fig. 4}(a) and Fig.\ref{Fig. 1}(a) we found that the maximum value of the momentum spectrum in two-color chirp-free laser pulse is two orders of magnitude higher than that of the one-color chirp-free laser pulse.

\begin{figure}
\begin{center}
\includegraphics*[width=12cm,keepaspectratio]{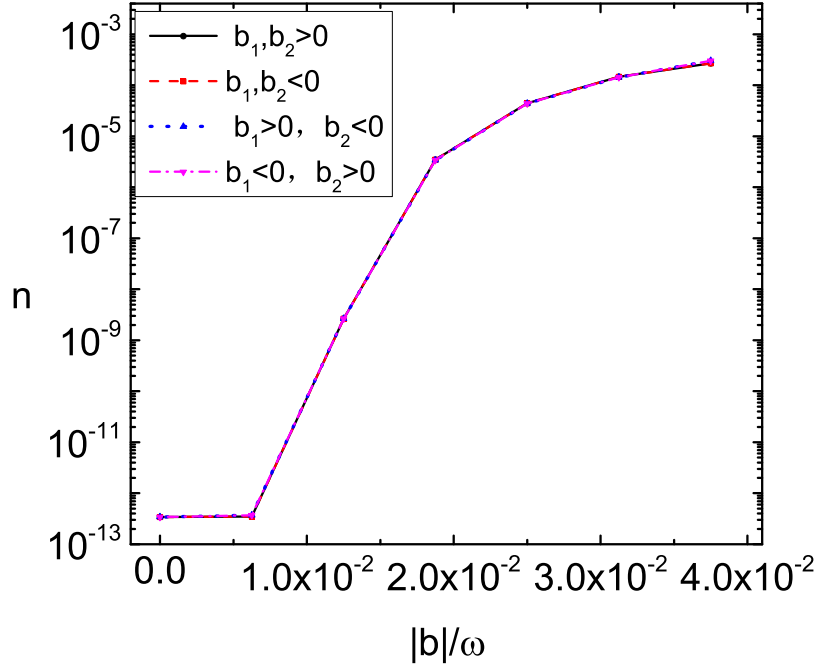}
\caption{\label{Fig. 5} (color online) Electron-positron number density vs the ratio of magnitude of frequency chirp to the original frequency, where $|b|/\omega=|b_1|/\omega_1=b_2/\omega_2$, in two-color pulse laser field $E_{1}(t)+E_{2}(t)$.}
\end{center}
\end{figure}

\begin{figure}
\begin{center}
\includegraphics*[width=12cm,keepaspectratio]{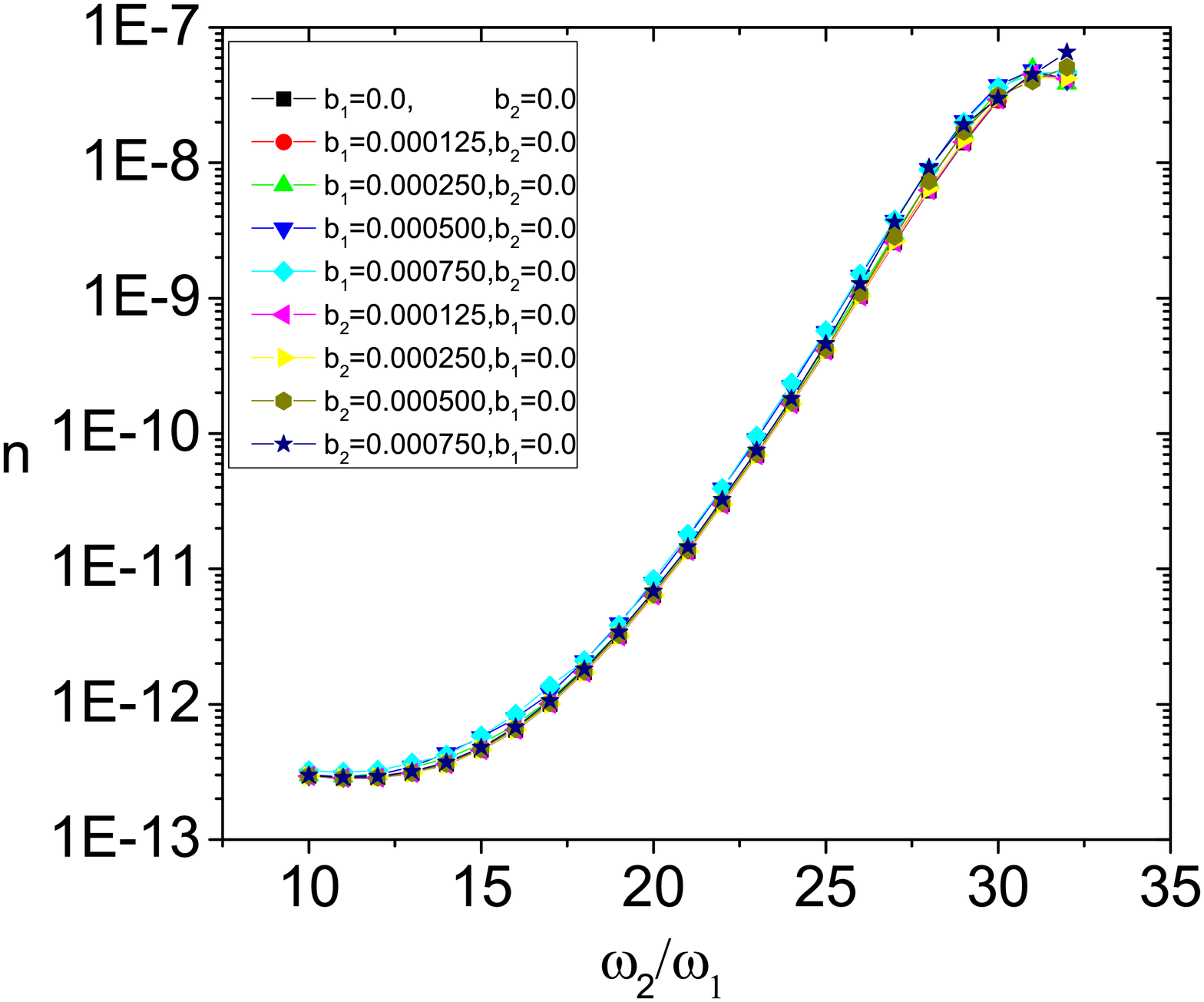}
\caption{\label{Fig. 6}(color online) Electron-positron number density vs the original two-frequency ratio in two-color pulse laser field $E_{1}(t)+E_{2}(t)$ with a single small frequency chirp of either $b_1$ or $b_2$.}
\end{center}
\end{figure}

In the case of two-color laser pulse first we study the effect of frequency chirping with the same signs of $b_{1}$ and $b_{2}$. The momentum spectrum curve is depicted in Fig.\ref{Fig. 4}(a) when $b_{1}=0.000125$, $b_{2}=10 b_{1}=0.00125$ (black solid line) and $b_{1}=-0.000125$, $b_{2}=10 b_{1}=-0.00125$ (red dotted line). The same as Fig.\ref{Fig. 4}(a) is shown in Fig.\ref{Fig. 4}(b) except that the frequency parameters are increased as $b_{1}=0.00075$, $b_{2}=10b_{1}=0.0075$ (black solid line) and $b_{1}=-0.00075$, $b_{2}=10b_{1}=-0.0075$ (red dotted line). Some phenomena can be observed from Figs.\ref{Fig. 4}(a) and (b). It is seen from Fig.\ref{Fig. 4}(a) that, by adding a small frequency chirp to the two-color laser pulse, the momentum spectrum shifts along the longitudinal momentum axis, especially for the positive frequency chirping. In comparison of chirping with chirp-free the maximum value of momentum spectrum does not increase remarkably but an oscillation structure with small peak appears. In the comparison of Fig.\ref{Fig. 4}(b) with Fig.\ref{Fig. 4}(a), when the chirp increases $6$ times, the maximum of the momentum spectrum increases $8$ orders of magnitude.

Second let us to examine the effect of frequency chirping with the opposite signs of $b_{1}$ and $b_{2}$ when $b_{1}>0$, $b_{2}=-10b_1<0$ and $b_{1}<0$, $b_{2}=10b_1>0$. The momentum spectrum are depicted in Fig.\ref{Fig. 4}(c) and Fig.\ref{Fig. 4}(d), where $|b_1|=0.000125$ and $|b_1|=0.00075$, respectively. Similarly after adding a small frequency chirp to the two-color laser pulse, the momentum spectrum shifts along the longitudinal momentum axis and the maximum value has no obvious increase but the appeared small peak lack of oscillation structure. On the other hand, when the chirp increases $6$ times the maximum of the momentum spectrum increases $9$ orders of magnitude. By the way in all cases of chirping the black solid line and the red dotted line has a  momentum-reverse symmetry except a nonzero momentum value as the symmetry point.

When $\omega_{1}=0.02$ and $\omega_{2}=0.2$ are fixed, by changing the $b_{1}$ and $b_{2}$, we get how the number density curve depends on $|b|/\omega$ in Fig.\ref{Fig. 5}. Since we keep $b_2/b_1=\omega_2/\omega_1$ so the horizonal axis can be characterized by a unified quantity $|b|/\omega$. It can be seen from the figure that, although the shape or value of momentum spectrum of the four fields are different but the number density of them are the same. It reveals the intrinsic symmetry again about the effect of chirping on momentum spectrum and concludes that this time-reverse or/and momentum-reverse symmetry leads to the same pair production.

\begin{figure}
\begin{center}
\includegraphics*[width=12cm,keepaspectratio]{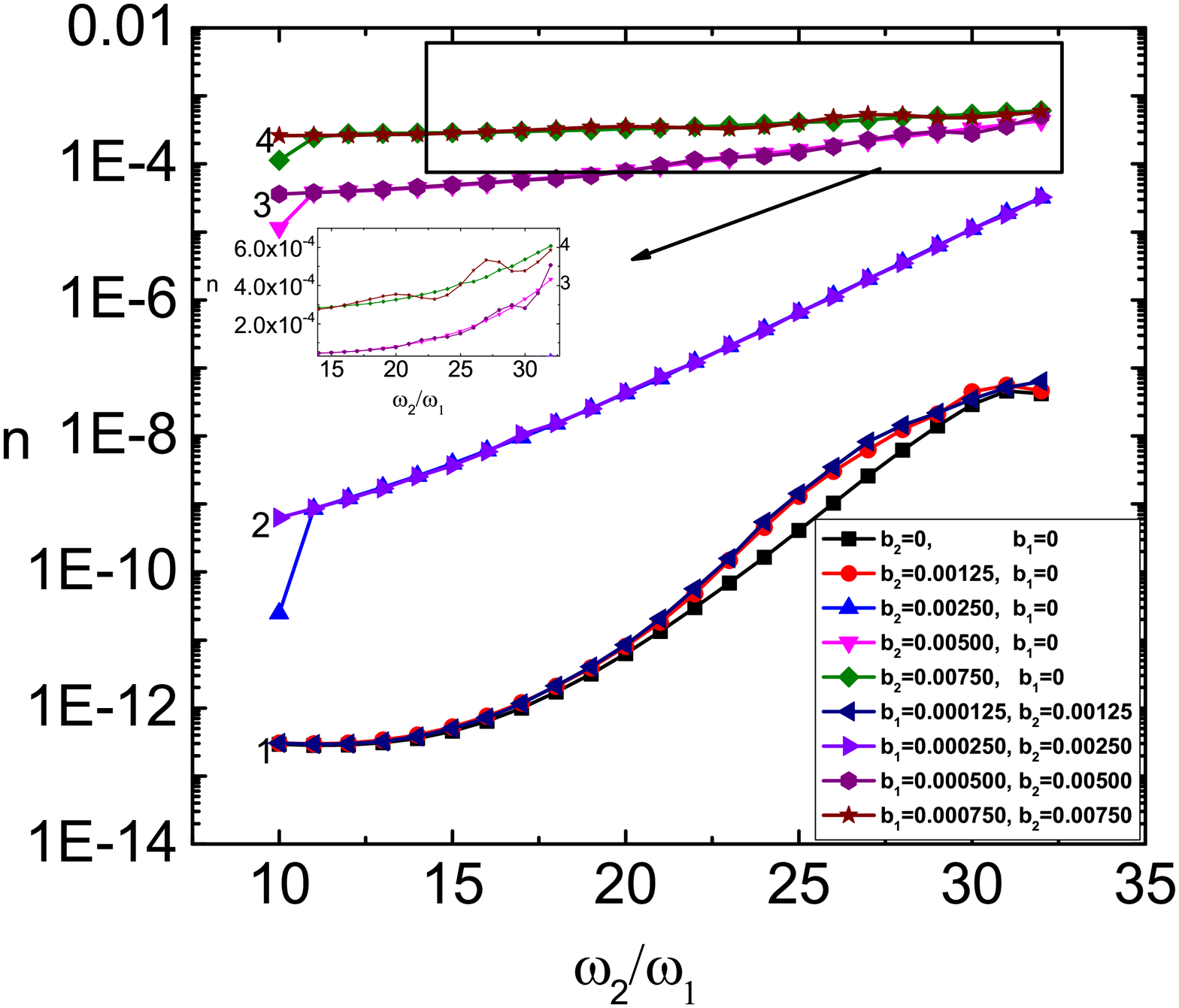}
\caption{\label{Fig. 7}(color online) Electron-positron number density vs the original two-frequency ratio in two-color pulse laser field $E_{1}(t)+E_{2}(t)$ with a relative large chirping for $b_2$ and the chirping of $b_1$ is either absent or small still.}
\end{center}
\end{figure}

Now we will investigate which laser pulse chirping field contributes more to the pair production rate when $\omega_{2}/\omega_{1}$ changes. We can see the effect of single frequency chirping, for example, keep $b_{2}=0$ fixed but change $b_{1}$, or keep $b_{1}=0$ fixed but change $b_{2}$. Some typical results about how the number density curves depend on the ratio $\omega_{2}/\omega_{1}$ for different single frequency chirping are shown in Fig.\ref{Fig. 6}. The concrete different values of $b_{1}$ and $b_{2}$ can be seen directly from the figure labels. As a comparison we plotted the line with black square symbol as the number density curve related to the chirp-free two-color laser pulse field. It is found that, as the ratio $\omega_{2}/\omega_{1}$ increases the number density increases slowly and tends to reach a saturation value. Different from Fig. \ref{Fig. 2}, in two-color laser pulse here, the number density curve does not exhibit the oscillation behavior. On the other hand it is interesting to find that the number density curves are in coincidence almost. It can conclude that in the two-color laser pulse field, if we add a small frequency chirp in either of $E_{1}(t)$ or $E_{2}(t)$, the effect of them on the pair production are the same and the increasing of the number density are not obvious compared to that of chirp-free one. The main reason of the number density increase is the increase of original frequency ratio of the two-color laser pulse field.

At last, we keep $b_{1}=0$ fixed and change a single parameter $b_{2}$, or we change $b_{1}$ within a small value but keep to change the corresponding $b_{2}=10b_{1}$ within a relatively large value. We depict the number density curves vs the ratio $\omega_{2}/\omega_{1}$ in Fig.\ref{Fig. 7} for different values of $b_{1}$ and $b_{2}$, which can be seen directly from the figure labels. In Fig.\ref{Fig. 7}, the value of $b_{2}$ are different in the four group of curves labeled $\rm{1}$, $\rm{2}$, $\rm{3}$ and $\rm{4}$ in the left side of curves. For example, the marked $1$ curves (except the chirp-free black solid line) correspond $b_{2}=0.00125$ and the marked $\rm{4}$ curves correspond $b_{2}=0.0075$. From these four sets of curves it is found that when the frequency ratio $\omega_{2}/\omega_{1}$ is not large, the number density curves seems very sensitive to the frequency chirp increase. For example, in the case of $\omega_{2}/\omega_{1}=10$, when the chirp parameter $b_2$ increases $6$ times from marked $\rm{1}$ to $\rm{4}$ the number density increase from $\sim 10^{-13}$ to $\sim 10^{-3}$, i.e. about $10$ orders of magnitude. As the frequency ratio $\omega_{2}/\omega_{1}$ increases the differences between the four groups of number density curves are reduced, but there are still $5$ orders of magnitude difference at $\omega_{2}/\omega_{1}\sim 30$. On the  other hand two curves in each group has little difference. But for the frequency ratio $\omega_{2}/\omega_{1}=10$, the difference of two curves is evident because whether of the small frequency chirp is added plays a role, especially in the case of group "2".

If we choose the number density curve of chirp-free, $b_{1}=0$ and $b_{2}=0$, as a reference, it is found that the differences in the curves $1$ are very small. There exist a little difference when the ratio $\omega_{2}/\omega_{1}$ increase. It may be caused by the diversification of the frequency \cite{Akal2014Electron,Eric2012Ramsey,Lindner2005Attosecond}. All the curves in each group $1$, $2$, $3$, $4$ are almost the same except when $\omega_{2}/\omega_{1}=10$. This is because that as the frequency ratio increased the small frequency chirp in the field $E_{1}(t)$ can be ignored. When the frequency chirp becomes larger, as the frequency chirp increased, the number density curves become less sensitive to the increasing of the frequency ratio $\omega_{2}/\omega_{1}$ , in the group $4$ the number density curves even don't change almost. By comparing the values of the curves, we found that the large frequency chirp of high frequency weak field $E_{2}(t)$ contributes more to increase the number density. Certainly to meet the condition $b<\omega/\tau$ we don't add too large frequency chirp parameters about $b_1$ as well $b_)2$. As the frequency ratio becomes larger the number density curve presents a small oscillation behavior, see the inset of Fig.\ref{Fig. 7}, this may be caused by the larger frequency chirp.

\section{Conclusions}

By solving the quantum Vlasov equation, we investigate the effect of frequency chirping  parameter on electron-positron pair production in one- and two-color laser pulse fields. The main findings of present study are the follows:

1. A positive or negative frequency chirp parameter is added to the one-or two-color pulse field, then the momentum spectrum and the number density curve of the created pairs in this field is obtained. It is found that in either one- or two-color laser pulse field the small frequency chirp shifts the momentum spectrum along the momentum axis, especially for the positive chirping. It can expand the detection probability by spectrometer.

2. Adding frequency chirp to the one-colo pulse field results in a smoothing number density curve and the widening of the peaks in multiphoton pair production process. The positive and negative frequency chirp parameters play the same role in increasing the number density. It is a natural result of time-reverse symmetry.

3. The sign change of frequency chirp parameter at the moment $t=0$ leads pulse shape and momentum spectrum to be symmetric and the number density to be increased. The chirping of first positive and then negative has a relative higher number density than that of first negative and then positive.

4. In the two-color pulse field, the number density is much higher than that in the one-color pulse field. The larger frequency chirped pulse field contributes more to increase the pair production rate.

5. In the two-color pulse field, the relation between the frequency ratio of two colors and the number density is not sensitive to the parameters of small frequency chirp added in either low frequency field or high frequency field but sensitive to the parameters of large frequency chirp added in high frequency field.

6. When the frequency chirp parameter increases $6$ times, the number density can increase $10$ orders of magnitude. As the original frequency ratio becomes larger the number density curve presents a small oscillation behavior.

We believe that the results obtained here is helpful to deepen the understanding of the pair production by including the effect of frequency chirping. And in an alternative way one can expect to control the pair production through the appropriate frequency chirping.

\begin{acknowledgments}
This work was supported by the National Natural Science Foundation of China (NSFC) under Grant no. 11475026 and no. 11175023.
\end{acknowledgments}

\end{document}